\documentclass{iaus}
\pdfoutput=0

\usepackage{graphicx}
\usepackage{subfig}
\usepackage{sidecap}
\usepackage{natbib}
\usepackage{amsmath}
\usepackage{MyStyle}

\title{The role of accretion disks in the formation of massive stars}

\author[R.~Kuiper, H.~Klahr, H.~Beuther \& Th.~Henning]   
{R.~Kuiper$^{1,2}$, H.~Klahr$^2$, H.~Beuther$^2$ \& Th.~Henning$^2$}

\affiliation{
$^1$
Argelander-Institut f\"ur Astronomie,
Rheinische Friedrich-Wilhelms-Universit\"at Bonn,\\
Auf dem H\"ugel 71,
D-53121 Bonn,
Germany \\
email: {\tt kuiper@astro.uni-bonn.de} \\
[\affilskip]
$^2$
Max-Planck-Institut f\"ur Astronomie,
K\"onigstuhl 17,
D-68169 Heidelberg,
Germany
}
\pubyear{2010}
\volume{270}  
\pagerange{}
\jname{Computational Star Formation}
\editors{J. Alves, B. Elmegreen, J. Girart \& V. Trimble}
\begin{document}

\maketitle

\begin{abstract}
We present radiation hydrodynamics simulations of the collapse of massive pre-stellar cores.
We treat frequency dependent radiative feedback from stellar evolution and accretion luminosity at a numerical resolution down to 1.27~AU.
In the 2D approximation of axially symmetric simulations, it is possible for the first time to simulate the whole accretion phase of several $10^5$~yr for the forming massive star and to perform a comprehensive scan of the parameter space.
Our simulation series show evidently the necessity to incorporate the dust sublimation front to preserve the high shielding property of massive accretion disks.
Our disk accretion models show a persistent high anisotropy of the corresponding thermal radiation field, yielding to the growth of the highest-mass stars ever formed in multi-dimensional radiation hydrodynamics simulations.
Non-axially symmetric effects are not necessary to sustain accretion.
The radiation pressure launches a stable bipolar outflow, which grows in angle with time as presumed from observations.
For an initial mass of the pre-stellar host core of 60, 120, 240, and 480\Msol the masses of the final stars formed in our simulations add up to
28.2, 56.5, 92.6, and at least 137.2\Msol respectively.
\end{abstract}

\firstsection 
\section{Introduction}
The understanding of massive stars suffers from the lack of a generally accepted formation scenario.
If the formation of high-mass stars is treated as a scaled-up version of low-mass star formation, a special feature of these high-mass proto-stars is the interaction of the accretion flow with the strong irradiation emitted by the newborn stars due to their short Kelvin-Helmholtz contraction timescale \citep{Shu:1987p1616}. 
Early one-dimensional studies 
\citep[e.g.][]{Larson:1971p1210, Kahn:1974p1200, Yorke:1977p1358} 
agree on the fact that the growing radiation pressure potentially stops and reverts the accretion flow onto a massive star.
But this radiative impact strongly depends on the geometry of the stellar environment \citep{Nakano:1989p1267}.
The possibility was suggested to overcome this radiation pressure barrier via the formation of a long-living massive circumstellar disk, which forces the generation of a strong anisotropic feature of the thermal radiation field.
Earlier investigations by \citet{Yorke:2002p1} tried to identify such an anisotropy, but
their simulations show an early end of the disk accretion phase shortly after its formation due to strong radiation pressure feedback.

\section{Method}
\subsection{The aim}
Aim of this study is to reveal the details of the radiation-dust interaction in the vicinity of the most massive star formed during the collapse of a pre-stellar core (gravitationally unstable fragment of a molecular cloud).
Due to this strong focus onto the core center, the physics in the outer core region have to be simplified, e.g.~the usage of quiescent initial conditions suppresses further fragmentation of the pre-stellar core.

\subsection{The code}
\label{sect:Physics}
For this purpose, we use our newly developed self-gravity radiation hydrodynamics code.
The evolution of the gas density, velocity, pressure, and total energy density is computed using the magneto-hydrodynamics code Pluto3 \citep{Mignone:2007p3421}, including full tensor viscosity.
The derivation and numerical details of the newly developed frequency dependent hybrid radiation transport method are summarized by \citet{Kuiper:2010p12874}.
Our implementation of Poisson's equation as well as the description of the dust and stellar evolution model are given in \citet{Kuiper2010b}.

The simulations are performed on a time independent grid in spherical coordinates.
The radially inner and outer boundary of the computational domain are semi-permeable walls, i.e.~the gas can leave but not enter the domain.
The resolution of the non-uniform grid is chosen to be 
$\left(\Delta r \mbox{ x } r~\Delta{\theta}\right)_\mathrm{min} = 1.27 \mbox { AU x } 1.04 \mbox{ AU}$
around the forming massive star and decreases logarithmically in the radial outward direction.
The accurate size of the inner sink cell is determined in a parameter scan presented in Sect.~\ref{sect:2Drmin}.

\subsection{Initial Conditions}
\label{sect:InitialConditions}
Our basic initial condition is very similar to the one used by \citet{Yorke:2002p1}.
We start from a cold ($T_0 = 20 \mbox{ K}$) pre-stellar core of gas and dust.
The initial dust to gas mass ratio is chosen to be $M_\mathrm{dust} / M_\mathrm{gas} = 1\%$.
The model describes a so-called quiescent collapse scenario without turbulent motion ($\vec{u}_r = \vec{u}_\theta = 0$).
The core is initially in slow rigid rotation $\left(|\vec{u}_\phi| / R = \Omega_0 = 5*10^{-13} \mbox{ Hz}\right)$.
The initial density slope drops with $r^{-2}$ and the outer radius of the cores is fixed to $r_\mathrm{max} = 0.1$~pc. 
The total mass $M_\mathrm{core}$ varies in the simulations from 60 up to 480 $\mbox{M}_\odot$. 

\section{Results}
\subsection{The dust sublimation front}
\label{sect:2Drmin}
\begin{figure}[bthp]
\begin{center}
\includegraphics[width=0.49\FigureWidth]{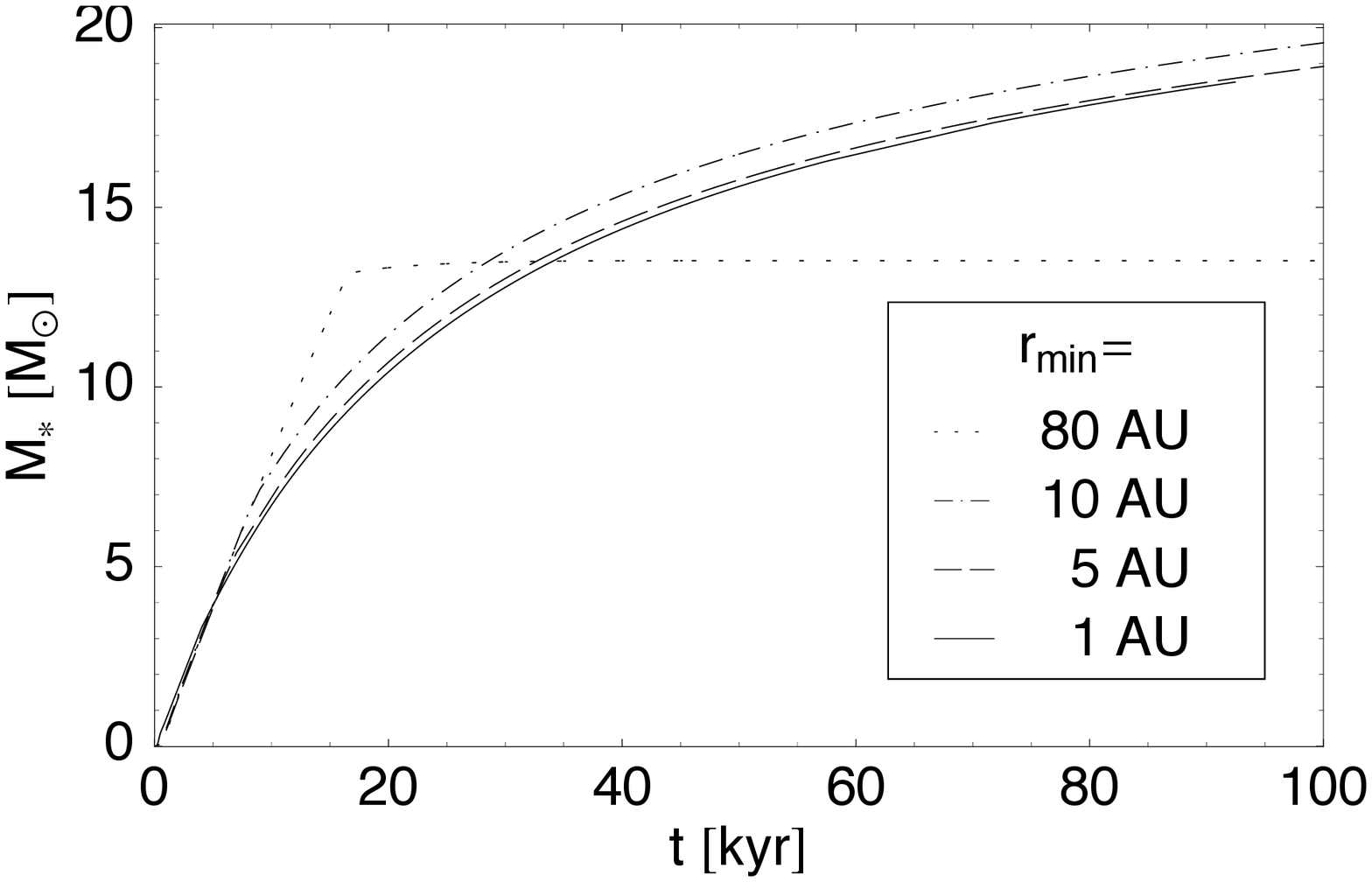} 
\includegraphics[width=0.51\FigureWidth]{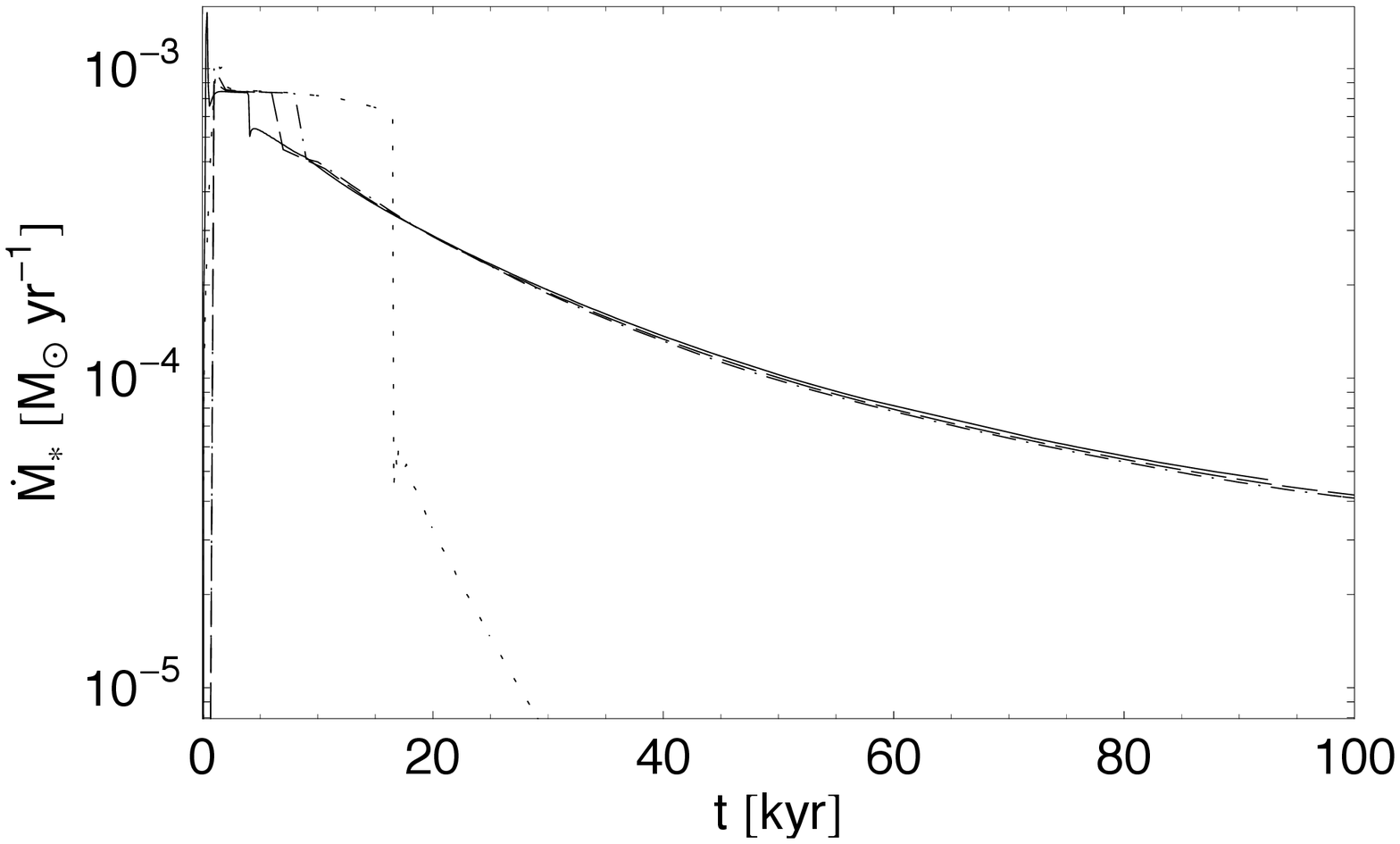} 
 \caption{Stellar mass $M_*$ (left panel) and accretion rate $\dot{M}_{*}$ (right panel) as a function of time $t$ for different radii $r_\mathrm{min}$ of the central sink cell 
 during a 60\Msol core collapse.}
 \label{fig:2D_Rmin_60Msol}
\end{center}
\end{figure}

In the following, we 
check the dependency of the stellar accretion rate on the radius $r_\mathrm{min}$ of the inner sink cell in four simulations with $r_\mathrm{min} =$~1,~5,~10,~and~80~AU.
We follow the long-term evolution of the runs for at least $10^5$~yrs.
The results are displayed in Fig.~\ref{fig:2D_Rmin_60Msol}.

The chosen location $r_\mathrm{min}$ of the inner boundary of the computational domain influences the resulting accretion rate in two distinguishable effects:
First, a smaller sink cell leads to a shorter free fall epoch, i.e.~to an earlier onset of the disk formation phase.
This effect of the artificial inner cut-off of the gas disk results on one hand in an overestimation of the final mass of the central star by approximately 1\Msol or below (left panel of Fig.~\ref{fig:2D_Rmin_60Msol}), but on the other hand does not influence the disk accretion epoch (right panel of Fig.~\ref{fig:2D_Rmin_60Msol}).

The second effect is related to a potential cut-off of the inner dust disk.
The region in the vicinity of the forming massive star will be heated up to temperatures beyond the dust sublimation temperature and a gap is formed between the central star and the dust disk.
For an inner sink cell radius $r_\mathrm{min}$ larger than the dust sublimation radius $r_\mathrm{subl}$, the region of radiative feedback is artificially shifted to higher radii including a strong decrease in density, opacity, and gravity.
As a result, the dust disk looses its shielding property, the thermal radiation field retains in major parts its isotropic character and the radiation pressure therefore stops the emerging disk accretion phase (cp.~the case of $r_\mathrm{min} = 80$~AU in Fig.~\ref{fig:2D_Rmin_60Msol}).
This dependency of the radiation pressure on the radius of the sink cell explains also the abrupt end of the accretion phase in the simulations by \citet{Yorke:2002p1},
who presented simulations of collapsing pre-stellar cores of $M_\mathrm{core} = 30 \Msol$, $60~\mbox{M}_\odot$ and $120 \mbox{ M}_\odot$ with a radius of the inner sink cell chosen 
to be 40, 80, and 160~AU respectively. 
Our subsequent simulations meet this concern by using an adequate central sink cell radius of $r_\mathrm{min} = 10$~AU. 


\subsection{The radiation pressure barrier}
\label{sect:2DMcore}
\begin{figure}[hbtp]
\begin{center}
 \includegraphics[width=0.6\FigureWidth]{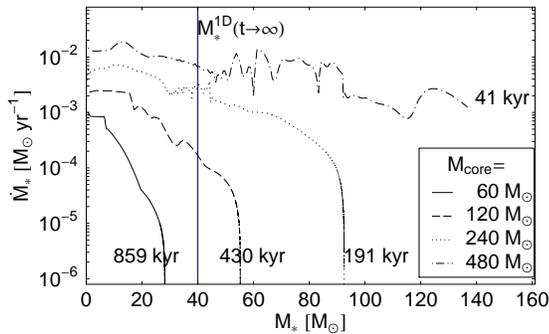} 
 \caption{
 Accretion rate $\dot{M}_*$ as a function of the actual stellar mass $M_*$ for four different initial core masses.
Also the periods of accretion are mentioned for each run.
The vertical line marks the upper mass limit found in the spherically symmetric accretion models. 
}
\label{fig:2D_McoreScan}
\end{center}
\end{figure}

The spherically symmetric 
accretion flow simulations
yield a maximum stellar mass of less than $40 \mbox{ M}_\odot$ independent of the initial core mass $M_\mathrm{core} \geq 60\Msol$ due to radiative feedback
\citep{Kuiper2010b}. 
We attack this radiation pressure barrier in
axially and midplane symmetric circumstellar disk geometry now.
Resulting accretion histories as a function of the actual stellar mass for different initial core masses of $M_\mathrm{core} = 60 \mbox{ M}_\odot$, $120 \mbox{ M}_\odot$, $240 \mbox{ M}_\odot$, and $480 \mbox{ M}_\odot$ are displayed in Fig.~\ref{fig:2D_McoreScan}.
In face of the additional centrifugal forces, the disk accretion easily breaks  through the upper mass limit of the final star of $M_*^\mathrm{1D} < 40 \mbox{ M}_\odot$ found in the spherically symmetric accretion models!
No upper mass limit of the final star is detected so far, but the star formation efficiency declines for higher mass cores. 
The depletion of the envelope by radiative forces decreases the large-scale accretion onto the midplane. 
The disk looses its shielding property and the radiation pressure starts to accelerate the remnant material in the outward direction.

\section{Summary}
\label{sect:summary}
We performed high-resolution radiation hydrodynamics simulations of monolithic pre-stellar core collapses including frequency dependent radiative feedback.
The dust sublimation front 
of the forming star could be resolved down to 1.27~AU.
The whole accretion phase of several $10^5$~yrs was computed.
The frequency dependent ray-tracing of our newly developed radiation module denotes the most realistic radiation transport method used in multi-dimensional hydrodynamic simulations of massive star formation by now. 

The broad parameter studies, especially regarding the size of the sink cell and the initial core mass, reveal new insights of the radiative feedback onto the accretion flow during the formation of a massive star:
In the case of disk accretion, the thermal radiation field generates a strong anisotropic feature.
We found that it is strict necessary to include the dust sublimation front in the computational domain to reveal the persistent anisotropy during the long-term evolution of the accretion disk.
The short accretion phases of the disks in the simulations by \citet{Yorke:2002p1} are a result of the fact that they did not include the dust sublimation front in their simulations, as clearly shown in our result of the parameter scan of the size of the central sink cell (see Sect.~\ref{sect:2Drmin}).
Additional feeding of the disk by unstable outflow regions due to the so-called ``3D radiative Rayleigh-Taylor instability'', as proposed in \citet{Krumholz:2009p10975}, would enhance this anisotropy but is not necessary.
In fact, preliminary analyses of our ongoing three-dimensional collapse simulations identify evolving gravitational torques in the massive circumstellar disk as well as the launching of non-axially symmetric outflows.
The accretion rate, driven by the angular momentum transport of the gravitational torques, is more episodically compared to the axially-symmetric runs, but results in a similar mean accretion rate.
The radiation pressure driven ouflow remains stable.


Finally, the central stars in our simulations of the disk accretion scenario grow far beyond the upper mass limit found in the case of spherically symmetric accretion flows. 
For an initial mass of the pre-stellar host core of 60, 120, 240, and 480\Msol the masses of the final stars formed  add up to
28.2, 56.5, 92.6, and at least 137.2\Msol respectively.
Indeed, the final massive stars are the most massive stars ever formed in a multi-dimensional radiation hydrodynamics simulation so far. 

\bibliographystyle{aa}
\bibliography{Papers}

\providecommand{\noopsort}[1]{}
\begin{thebibliography}{10}
\expandafter\ifx\csname natexlab\endcsname\relax\def\natexlab#1{#1}\fi

\bibitem[{Kahn(1974)}]{Kahn:1974p1200}
Kahn, F.~D. 1974, A{\&}A, 37, 149

\bibitem[{Krumholz {et~al.}(2009)Krumholz, Klein, McKee, Offner, \&
  Cunningham}]{Krumholz:2009p10975}
Krumholz, M.~R., Klein, R.~I., McKee, C.~F., Offner, S. S.~R., \& Cunningham,
  A.~J. 2009, Science, 323, 754

\bibitem[{Kuiper {et~al.}(2010{\natexlab{a}})Kuiper, Klahr, Dullemond, Kley, \&
  Henning}]{Kuiper:2010p12874}
Kuiper, R., Klahr, H., Dullemond, C., Kley, W., \& Henning, T.
  2010{\natexlab{a}}, A{\&}A, 511, 81

\bibitem[{Kuiper {et~al.}(2010{\natexlab{b}})Kuiper, \noopsort{Z Z}{Klahr, H.},
  Beuther, \& Henning}]{Kuiper2010b}
Kuiper, R., \noopsort{Z Z}{Klahr, H.}, Beuther, H., \& Henning, T.
  2010{\natexlab{b}}, ApJ, in press

\bibitem[{Larson \& Starrfield(1971)}]{Larson:1971p1210}
Larson, R.~B. \& Starrfield, S. 1971, A{\&}A, 13, 190

\bibitem[{Mignone {et~al.}(2007)Mignone, Bodo, Massaglia, Matsakos, Tesileanu,
  Zanni, \& Ferrari}]{Mignone:2007p3421}
Mignone, A., Bodo, G., Massaglia, S., {et~al.} 2007, ApJS, 170, 228

\bibitem[{Nakano(1989)}]{Nakano:1989p1267}
Nakano, T. 1989, ApJ, 345, 464

\bibitem[{Shu {et~al.}(1987)Shu, Lizano, \& Adams}]{Shu:1987p1616}
Shu, F.~H., Lizano, S., \& Adams, F.~C. 1987, in: Star forming regions, ed. M.
  Peimbert, J. Jugaku, 115, 417

\bibitem[{Yorke \& Kr{\"u}gel(1977)}]{Yorke:1977p1358}
Yorke, H.~W. \& Kr{\"u}gel, E. 1977, A{\&}A, 54, 183

\bibitem[{Yorke \& Sonnhalter(2002)}]{Yorke:2002p1}
Yorke, H.~W. \& Sonnhalter, C. 2002, ApJ, 569, 846

\end{thebibliography}
\end{document}